# Improving positioning accuracy of the mobile laser scanning in GPS-denied environments: An experimental case study

WI. Liu, Zhixiong Li, Shuaishuai Sun, Reza Malekian, Zhenjun Ma, Weihua Li

*Abstract*—The positioning accuracy of the mobile laser scanning (MLS) system can reach the level of centimeter under the conditions where GPS works normally. However, in GPS-denied environments this accuracy can be reduced to the decimeter or even the meter level because the observation mode errors and the boresight alignment errors of MLS cannot be calibrated or corrected by the GPS signal. To bridge this research gap, this paper proposes a novel technique that appropriately incorporates the robust weight total least squares (RWTLS) and the full information maximum likelihood optimal estimation (FIMLOE) to improve the positioning accuracy of the MLS system under GPS-denied environment. First of all, the coordinate transformation relationship and the observation parameters vector of MLS system are established. Secondly, the RWTLS algorithm is used to correct the 3D point observation model; then the uncertainty propagation parameter vector and the boresight alignment errors between the laser scanner frame and the IMU frame are calibrated by FIMLOE. Lastly, experimental investigation in indoor scenarios was performed to evaluate the effectiveness of the proposed method. The experimental results demonstrate that the proposed method is able to significantly improve the positioning accuracy of an MLS system in GPS-denied environments.

Manuscript received #########, accepted #########, date of publication #########, date of current version #########. This work is supported by National Key R&D Program of China (2018YFC0604503), National Natural Science Foundation of China (U1610251), Applied Fundamental Research project of Qingdao (2019-9-1-14-jch), Taishan Scholar (tsqn201812025), Australia ARC DECRA (No. DE190100931) and the Priority Academic Program Development (PAPD) of Jiangsu Higher Education Institutions.

Dr. WI Liu is with the School of Mechanical and Electrical Engineering, China University of Mining and Technology, Xuzhou 221116, China and Jiangsu Collaborative Innovation Center of Intelligent Mining Equipment, China University of Mining and Technology, Xuzhou 210008, China; (e-mail: 4830@cumt.edu.cn)

Dr. Zhixiong Li is with School of Engineering, Ocean University of China, Tsingdao 266100,China; and School of Mechanical, Materials, Mechatronic and Biomedical Engineering, University of Wollongong, Wollongong, NSW 2522, Australia; (email: zhixiong.li@ieee.org)

Dr. Shuaishuai Sun is with the New Industry Creation Hatchery Center, Tohoku University, Sendai 980-8577, Japan (e-mail: shuaishuai.sun.b1@tohoku.ac.jp)

Dr. R. Malekian is with Department of Computer Science and Media Technology, Malmö University, Malmö, 20506, Sweden (email: reza.malekian@ieee.org)

A. Professor Zhenjun Ma and Prof. Weihua Li are with School of Mechanical, Materials, Mechatronic and Biomedical Engineering, University of Wollongong, Wollongong, NSW 2522, Australia; (email: zhenjun@uow.edu.au; weihuali@uow.edu.au).

Corresponding author: Z. Li (zhixiong.li@ieee.org)

## I. INTRODUCTION

The mobile laser scanning (MLS) system is a kinematic platform and mainly consists of laser scanner, inertial measurement unit (IMU), GPS receiver as well as other ancillary devices on a moving platform; this system can be used to generate 3D point cloud data of the surrounding scene with high precision, convenience, efficiency and effectiveness. These data are useful for many applications such as 3D landscape modeling for visualization in planning, simulations for environmental management and navigation for robots and vehicles. For these practical applications, positioning accuracy is essentially important, particularly in GPS-denied environments where the position accuracy can be reduced to the decimeter or even the meter level due to the trajectory errors of the laser scanner and the IMU drift. Hence, it is crucial to improve the positioning accuracy of MLS in GPS-denied environments [1-3].

In GPS-denied environments, the main errors affecting the positioning accuracy in an MLS system can be categorized into three main sources [4-8]. The first one is laser scanner error. The errors in range and angular measurements of the time-of-flight laser scanner lead to the uncertainty in locating the actual positions of the scanned points. IMU attitude error is the second inaccuracy source. The principal role of IMU is to provide angular velocity observations which can be integrated into angular orientation information (roll, pitch, and heading) of the IMU body frame with respect to the MLS system local frame. Together with position data, it enables the point data in the MLS local frame to be transformed into the earth-centered-earth-fixed frame. Thus, all points in the point cloud are projected into a common reference frame and the IMU attitude data may affect every point. Therefore any uncompensated errors from the IMU will directly impact the geometric quality of the point cloud. The third source is boresight alignment error. The so-called *boresight angles* are the orientations of the laser scanner frame with respect to the IMU body frame and the lever arm offset. Because the boresight angles cannot be measured directly, they are obtained through a calibration process. There are inevitable residual adjustment errors present in the alignment estimates. The lever arm offset values can be obtained either through calibration or measurement. Noting that the required boresight angles are vector components, not just lengths, some realization of the relevant coordinate systems is necessary.

In order to improve the position accuracy and eliminate error sources for the MLS system in GPS-denied environments, previous researchers have made great progress on data-driven and model-driven techniques. The data-driven techniques can be directly used to correct the laser scanning point clouds data based the on ground control points and the different correction algorithms [9-10]. Mao et al. [11] proposed a least squares collocation (LSC) technique to increase the mobile LiDAR accuracy in GPS hostile environments. Shi [12] presented an adaptive mapped smooth fitting method on the basis of least squares support vector machine (LS-SVM) for digital surface model generation of airborne LiDAR scanning data. Gneeniss et al. [13] utilized a total least squares (TLS) surface matching algorithm to align a dense network of photogrammetric points to the LiDAR reference surface, allowing for the automatic extraction of LiDAR control points. Hans-Gerd [14] presented a formulation of least squares (LS) matching based on the original data points in a triangulated irregular network structure, avoiding the degrading effects caused by the interpolation. Lee et al. [15] solved the observation equations of the airborne laser scanning system using the LS method. A set of affine transformation equations were produced to adjust the horizontal error. Xi et al. [16] proposed the TLS method for active view registration of three-dimensional line laser scanning data. Furthermore, the simultaneous localization and mapping (SLAM) algorithms have attracted extensive attention in recent years. SLAM aims to build a map of the environment while simultaneously determine the position of a moving sensor platform (most notably in photogrammetry and computer vision) [17-19]. SLAM can use the nonlinear least squares (NLS) to correct the position information.

The model-driven techniques usually establish the mathematical models for the MLS system and analyze the error sources to calibrate the errors. Chen et al. [20] proposed a new method for boresight misalignment calibration of the charge-coupled device CCD camera in an airborne LiDAR system without ground control points. Darren et al. [21] used a planar patch least-squares approach to determine the boresight angles and the lever-arm offsets of an MLS, and calibrated the boresight MLS operation in a backpack mode to acquire multiple data sets in an area that contains dense ground control points. Habib et al. [22] provided a tool for the quality control of the LiDAR point cloud and analyzed possible systematic and random errors as well as their impact on the laser scanning surface. Ye et al. [23] proposed a calibration method with small rotation-angle corrections for the exterior orientations of a vehicle laser imaging system.

Although LS, LS-SVM, NLS, LSC and TLS can be used to resolve the problem of parameter estimation in laser scanning point clouds data, the estimated parameters and the position accuracy can be influenced when 3D point observation model and uncertainty propagation parameter vector of MLS system are contaminated by laser scanner errors and attitude errors of IMU. The main limitation of the existing methods is that they assume that the trajectory errors and IMU drift errors in the MLS system are very small or negligible. If this assumption is not available or sufficiently accurate, then the estimated parameters and position accuracy will significantly degrade [24, 25]. Literature review shows that the robust weight totals least squares (RWTLS) method is based on the robust estimation equivalent weight rule and the Newton-Gauss method, which utilizes standardized residual to construct the weight factor function and uses the median technique to acquire a variance component estimator. Therefore, RWTLS possesses good robustness in both 3D point observation model and uncertainty propagation parameter vector in MLS [26-28]. However, RWTLS cannot calibrate the boresight alignment errors of MLS. To solve this problem, the full information maximum likelihood optimal estimation (FIMLOE) can be integrated with RWTLS. The FIMLOE method uses an approximate Newton method to identify boresight alignment parameters between the IMU and the laser scanner, which enables FIMLOE to yield very low proportion of convergence failures and to provide near-optimal boresight alignment errors [29-31]. Hence, it is reasonable to develop a new RWTLS-FIMLOE method to improve MLS positioning accuracy in GPS-denied environments. However, to our best knowledge, RWTLS-FIMLOE has not been found in literature.

For this reason, this paper proposes a novel RWTLS-FIMLOE method to improve the MLS system positioning accuracy in GPS-denied environments. The contributions of this paper are: (1) the proposed RWTLS-FIMLOE method is introduced to correct the 3D point observation model and the uncertainty propagation parameter vector; it is also used to calibrate the boresight alignment errors between the laser scanner frame and the IMU frame; (2) an MLS mathematical model is established to analyze the effects of each individual error source on the positioning accuracy. Experimental results demonstrate that the proposed method achieved significant improvement in terms of the MLS positioning accuracy in GPS-denied environments.

## II. MODELING FOR MLS SYSTEM

In GPS-denied environments, the IMU provides the position and orientation for moving platform during MLS measuring process. The laser scanner measures the distances from the sensor to the scanned target point and records the rotating angles of the laser beam. Thus, the accuracy of the laser scanning point clouds depends on the quality in terms of reliability and accuracy of the moving platform trajectory. However, the IMU provides the MLS system a moving trajectory, instead of the fixed control points through a traditional control network. As a result, both of the IMU and the laser scanner direct referencing coordinate frame make up the total error budget together. Generally, an MLS system includes three basic coordinate frames. Figure 1 shows the relationship among these frames [8, 32]:

(1) The world coordinate frame $\{W\}$.

(2) The integrated moving platform and the IMU coordinate

frame $\{I\}$.

(3) The laser scanner coordinate frame $\{L\}$.

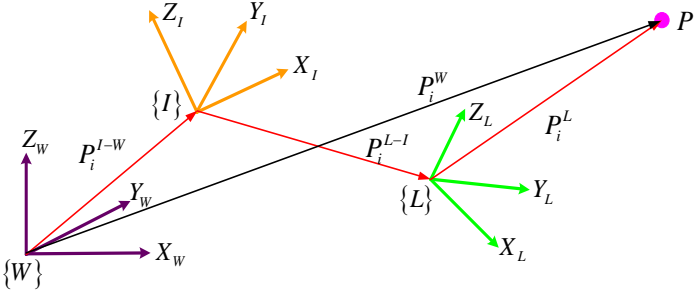

Fig. 1. Coordinate relationship between $\{W\}$, $\{I\}$ and $\{L\}$ frame

As shown in Figure 1, the $P_i^W$ is the coordinates of target point $P_i$ in the frame $\{W\}$, $P_i^L$ is the position of target point $P_i$ in the frame $\{L\}$, the $P_i^{I-W}$ is the relative position vector of the frame $\{I\}$ to the frame $\{W\}$, and the $P_i^{L-I}$ is the relative position vector of the frame $\{L\}$ to the frame $\{I\}$. The base equation of the Coordinate relationship between frames $\{W\}$, $\{I\}$ and $\{L\}$ can be expressed as:

$$P_i^W = R_{I-W} P_i^{I-W} + T_{I-W} \qquad (1)$$

$$P_i^{L-I} = R_{L-I}(\Omega,\varphi,\kappa) P_i^L + T_{L-I} \qquad (2)$$

where $R_{L-I}$ is the constant rotation matrix between frame $\{L\}$ and frame $\{I\}$. $\Omega,\varphi,\kappa$ are the boresight angles, $T_{L-I}$ is the translation matrix between frame $\{L\}$ and frame $\{I\}$.

The $T_{I-W}$ and $R_{I-W}$ are the translation matrix and rotation matrix between frame $\{I\}$ and frame $\{W\}$.

$$R_I^W = R_{NED}^{ENU} R(\theta_x) R(\theta_y) R(\theta_z) \qquad (3)$$

where

$$R(\theta_x) = \begin{bmatrix} 1 & 0 & 0 \\ 0 & \cos\theta_x & -\sin\theta_x \\ 0 & \sin\theta_x & \cos\theta_x \end{bmatrix}, \quad R(\theta_y) = \begin{bmatrix} \cos\theta_y & 0 & \sin\theta_y \\ 0 & 1 & 0 \\ -\sin\theta_y & 0 & \cos\theta_y \end{bmatrix},$$

$$R(\theta_z) = \begin{bmatrix} \cos\theta_z & -\sin\theta_z & 0 \\ \sin\theta_z & \cos\theta_z & 0 \\ 0 & 0 & 1 \end{bmatrix}$$ is the rotation matrix between the frame $\{I\}$ and frame $\{W\}$ (in the North-East-Down coordinate system) with $\theta_z, \theta_y, \theta_x$ which represent the roll, pitch and yaw Euler angles given by the IMU. $R_{NED}^{ENU} = \begin{pmatrix} 0 & 1 & 0 \\ 1 & 0 & 0 \\ 0 & 0 & -1 \end{pmatrix}$ is the constant rotation matrix between the North-East-Down coordinates and the East-North-Up coordinate frame.

The measuring model of $P_i^L$ is given by:

$$P_i^L = \begin{pmatrix} X_0^L \\ Y_0^L \\ Z_0^L \end{pmatrix} + \begin{pmatrix} \rho_i \cos\alpha_i \sin\beta_i \\ \rho_i \cos\alpha_i \cos\beta_i \\ \rho_i \sin\alpha_i \end{pmatrix} \qquad (4)$$

where $\begin{pmatrix} X_0^L & Y_0^L & Z_0^L \end{pmatrix}^T$ is the mirror center offset of laser scanner, $\rho$ is the measurement range, $\alpha$ and $\beta$ are the vertical angle and the horizontal angle of laser scanner, respectively.

According to the Equations (2)-(4), the Equation (1) can be rewritten as:

$$\begin{pmatrix} X_i^W \\ Y_i^W \\ Z_i^W \end{pmatrix} = R(\theta_x) R(\theta_y) R(\theta_z) R_{NED}^{ENU} \begin{pmatrix} R_L^I(\Omega,\varphi,\kappa) \begin{pmatrix} X_0^L \\ Y_0^L \\ Z_0^L \end{pmatrix} + \begin{pmatrix} \rho_i \cos\alpha_i \sin\beta_i \\ \rho_i \cos\alpha_i \cos\beta_i \\ \rho_i \sin\alpha_i \end{pmatrix} + \begin{pmatrix} X_{L-I} \\ Y_{L-I} \\ Z_{L-I} \end{pmatrix} \end{pmatrix} + \begin{pmatrix} X_{I-W} \\ Y_{I-W} \\ Z_{I-W} \end{pmatrix}$$

(5)

Equation (5) shows that the coordinates of points $P_i^W$ in the frame $\{W\}$ depends on the observation parameters vector:

$$\Phi = \left[\theta_x, \theta_y, \theta_z, \Omega, \varphi, \kappa, \rho, \alpha, \beta, X_0^L, Y_0^L, Z_0^L, X_{L-I}, Y_{L-I}, Z_{L-I}, X_{I-W}, Y_{I-W}, Z_{I-W}\right]^T$$

Then the RWTLS-FIMLOE integration technique is used to compute the uncertainty of $\Phi$ for every 3D point $P_i^W$.

## III. RWTLS-FIMLOE INTEGRATION METHOD

The block diagram of the proposed RWTLS-FIMLOE algorithm is illustrated in Figure 2. Firstly, the boresight alignment parameter errors between the laser scanner and IMU are calibrated using FIMLOE algorithm. According to Equation (5), the laser scanning points are transformed into the frame $\{W\}$. Then the trajectory errors are corrected, the laser scanning control targets coordinate is extracted, and the uncertainty propagation parameter vector is calculated. Lastly, the 3D point observation model and the uncertainty propagation parameter vector are corrected using RWTLS algorithm.

### A. RWTLS Correction for MLS Observation Model and Uncertainty Propagation Vector

According to Equation (5), the observation model of $P_i^W$ can be simplified as [21, 33]:

$$P = (A - E_A)\Phi + e_P \qquad (6)$$

where $e_P$ represents the random error vector of $P$, $A$ denotes the $m \times n$ coefficient matrix, $E_A$ denotes the random error matrix of $A$.

Supposing that $A$ is unstructured, then the random elements should be calculated firstly and the independent random variables from coefficient matrix are extracted, Equation (6) can be rewritten as:

$$P = \left(\Phi^T \otimes I_m\right)\left(h + B(a - e_a)\right) + e_P \qquad (7)$$

where $B$ represents the deterministic matrix, $h$ denotes the deterministic constant vector of $A$, $a$ represents the independent variables vector, $I_m$ represents the unit matrix,

$e_a$ is the random error vector for $a$. Therefore, the coefficient matrix $A$ is calculated as:
$$A = ivec(h + Ba) \quad (8)$$
where the "$ivec$" denotes the original matrix vector with $m \times n$ dimension.

The MLS stochastic model can be expressed as:
$$\begin{pmatrix} e_P \\ e_a \end{pmatrix} \sim \left( 0, \sigma_0^2 \begin{pmatrix} Q_P & 0 \\ 0 & Q_a \end{pmatrix} \right) \quad (9)$$
where $Q_a$ and $Q_P$ represent the positive definite cofactor matrix of $e_a$ and $e_P$, respectively.

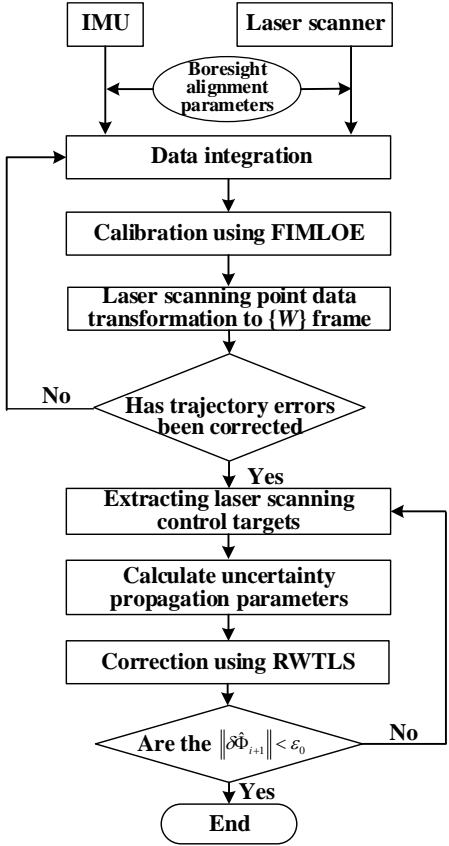

Fig. 2 Flowchart of RWTLS - FIMLOE algorithm.

Supposing that through $i-th$ iteration, the parameter estimator vectors $\Phi_i$ can be derived, the predictive residual vector of $a$ is $e_{a_i}$. The right-hand member of Equation (7) is expressed through Taylor series expansion at $(\Phi_i, e_{a_i})$:
$$\begin{aligned} P &= \left(\Phi_i^T \otimes I_m\right)(h + Ba) - \left(\Phi_i^T \otimes I_m\right) B e_a + \\ &\quad ivec\left(h + B(a - e_{a_i})\right) \delta \Phi + e_P \\ &= A\Phi_i + A_i \delta \Phi - \left(\Phi_i^T \otimes I_m\right) B e_a + e_P \end{aligned} \quad (10)$$
where $A_i = ivec\left(h + B(a - e_{a_i})\right) = A - ivec\left(B e_{a_i}\right)$.

According to the Equation (10), $\delta \Phi$ represents the vector $\Phi_i$ little correction values, the RWTLS Lagrange objective function can be calculated as:
$$\begin{aligned} f(e_P, e_a, \Phi, K) &= e_P^T Q_P^{-1} e_P + e_a^T Q_a^{-1} e_a + \\ &\quad 2K^T \left(P - e_P - A\Phi_i - A_i \delta \Phi + \left(\Phi_i^T \otimes I_m\right) B e_a\right) \end{aligned} \quad (11)$$

where $K$ is the Lagrange multipliers vector with the dimension $m \times 1$.

According to the Equation (11), the solution for the function $f$ can be computed as:
$$\left. \frac{\partial f}{\partial e_P} \right|_{\tilde{e}_P, \tilde{e}_a, \delta\hat{\Phi}, \hat{K}} = Q_P^{-1} \tilde{e}_P - \hat{K} = 0 \quad (12)$$

$$\left. \frac{\partial f}{\partial e_a} \right|_{\tilde{e}_P, \tilde{e}_a, \delta\hat{\Phi}, \hat{K}} = Q_a^{-1} \tilde{e}_a + B^T \left(\Phi_i \otimes I_m\right) \hat{K} = 0 \quad (13)$$

$$\left. \frac{\partial f}{\partial \delta \Phi} \right|_{\tilde{e}_P, \tilde{e}_a, \delta\hat{\Phi}, \hat{K}} = -A_i^T \hat{K} = 0 \quad (14)$$

$$\left. \frac{\partial f}{\partial K} \right|_{\tilde{e}_P, \tilde{e}_a, \delta\hat{\Phi}, \hat{K}} = P - \tilde{e}_P - \Phi_i - A_i \delta\hat{\Phi} + \left(\Phi_i^T \otimes I_m\right) B \tilde{e}_a = 0 \quad (15)$$

where "~" and "^" represent the predicted and the estimated, respectively.

According to the Equations (12)-(15), the following equations are obtained:
$$\delta \hat{\Phi}_{i+1} = \left(A_i^T Q_{c_i}^{-1} A_i\right)^{-1} A_i^T Q_{c_i}^{-1} \left(P - A\Phi_i\right) \quad (16)$$

$$\hat{\Phi}_{i+1} = \delta\hat{\Phi}_{i+1} + \Phi_i = \left(A_i^T Q_{c_i}^{-1} A_i\right)^{-1} A_i^T Q_{c_i}^{-1} \left(\left(\Phi_i^T \otimes I_m\right) B \tilde{e}_{a_i}\right) \quad (17)$$

$$\tilde{e}_{P_{i+1}} = \Phi_P Q_{c_i}^{-1} \left(P - A\Phi_i - A_i \delta\hat{\Phi}_{i+1}\right) \quad (18)$$

$$\tilde{e}_{a_{i+1}} = -Q_a B^T \left(\Phi_i \otimes I_m\right) Q_{c_i}^{-1} \left(P - A\Phi_i - A_i \delta\hat{\Phi}_{i+1}\right) \quad (19)$$

where $Q_{c_i} = Q_P + \left(\Phi_i^T \otimes I_m\right) B Q_a B^T \left(\Phi_i \otimes I_m\right)$.

The Equations (16)-(19) denote the iterative procedure of RWTLS algorithm, during this process, the threshold $\varepsilon_0$ should be given to terminate iteration when $\left\| \delta \hat{\Phi}_{i+1} \right\| < \varepsilon_0$, then the positive cofactor matrix $Q_a$ instead of $Q_A$ can be calculated, which will bring convenience in constructing RWTLS model.

### B. FIMLOE Calibration for MLS Boresight Alignment Errors

In Section A, the boresight alignment parameters between the frame $\{L\}$ and the frame $\{I\}$ are assumed to have been calibrated accurately in advance. However, it is not always the case; the calculating method of these parameters is an important problem in the MLS system. Therefore, the FIMLOE algorithm is used to calibrate the boresight alignment errors between the frame $\{L\}$ and the frame $\{I\}$.

According to Equation (5), the boresight alignment parameters vector is treated as time-invariant tendency variables denoted by $\Psi$, which can be collected as:
$$\Psi = \left[\Omega, \varphi, \kappa, \rho, \alpha, \beta, T_0^L, T_L^I\right]^T \quad (20)$$

Therefore, the FIMLOE algorithm is to solve the following optimization problem [34]:
$$P_{FIMLOE}^{L-I}, \Psi_{FIMLOE} = \arg\min_{P, \Psi} f(P, \Psi, U, V) \quad (21)$$

where $P_{FIMLOE}^{L-I}, \Psi_{FIMLOE}$ are the full information maximum likelihood estimation of the state position vector and the alignment parameters vector, respectively; $f(\cdot)$ is the joint probability density function; $U$ is the laser scanner measurement of each scanning point in the time instances $t$ at the frame $\{L\}$; $V$ is IMU measurement results in the time instances $t$ at the frame $\{I\}$.

In the MLS measuring process, the IMU noise process is dependent variable, the laser scanner noise process is independent, and the laser scanner measuring points depend on the value of $U$ in the time instance $t$. Therefore, the function $f(\cdot)$ can be written as:

$$f(P,\Psi,U,V) = \prod_{j=0}^{M}\prod_{t=0}^{T} f(U_{jt}|P,\Psi) \times \prod_{t=0}^{T} f(V_t|P,\Psi) \quad (22)$$

where $M$ is the total number of the laser scanning points, $T$ is the total measuring time; $U_{jt}$ is the state in the time instance $t$ and it lies on the total measurement points $M$; $V_t$ is the IMU measurements in the time instance $t$.

The calibration procedure of FIMLOE algorithm mainly includes the following:

(1) Nominal state position vector and alignment parameters vector

Although the measurement models of IMU and the estimation process are nonlinear, the nominal state position vector $\bar{P}$ and the alignment parameters vector $\bar{\Psi}$ are linearized, and they are perturbed with the state position error and the alignment parameters error $\delta P$, $\delta \Psi$:

$$\begin{cases} \bar{P} = P \otimes \delta P \\ \bar{\Psi} = \Psi \otimes \delta \Psi \end{cases} \quad (23)$$

(2) Minimal state position vector and alignment parameters vector representation

The transform from minimal states to nominal states can be represented using function $\breve{P}$ and $\breve{\Psi}$:

$$\begin{cases} \breve{P} = H(\breve{\Psi}) \cdot P + G(\bar{P})\delta U \delta V \\ \breve{\Psi} = \Psi + G(\bar{\Psi})\delta U \delta V \end{cases} \quad (24)$$

where $G$ is the driving white noise process on the minimal state; $H$ is the derivative of the alignment parameter function regarding to the minimal states function $\breve{P}$ and $\breve{\Psi}$.

(3) FIMLOE calibration

According to Equation (22) and Equation (24), the calibrated vector $P_{FIMLOE}^{L-I}, \Psi_{FIMLOE}$ using FIMLOE algorithm is equivalent to solve the optimization problems as following.

$$\begin{cases} F(P_{FIMLOE}^{L-I}) = \arg\min_{\breve{P}_{jt}} \left\| \sum_{j=0}^{M}\sum_{t=0}^{T} H(j|j-1) \cdot H(t|t-1) \cdot f(\breve{P}_{jt}) \right\|_{U,V}^2 \\ F(\Psi_{FIMLOE}) = \arg\min_{\breve{\Psi}_t} \left\| \sum_{t=0}^{T} H(t|t-1) \cdot f(\breve{\Psi}_t) \right\|_{U,V}^2 \end{cases}$$

$$(25)$$

where $\|\cdot\|^2$ is referred to the vector Euclidean length; $F(\cdot)$ is the FIMLOE minimal joint probability density function.

Therefore, the misalignment calibration error can be computed by Equation (25) when more than two controlling points are measured.

## IV. EXPERIMENTS

To verify the effectiveness of the proposed algorithm, a series of experiments in indoor and outdoor environments were conducted, as shown in Figure 3. The indoor experiments were performed on the fifth floor of a building in China University of Mining and Technology, and the outdoor experiments were carried out around the teaching building in the campus. The experiments were conducted by using a mobile laser scanning system manufactured by HiScan [35] (Figure 4), which consists of a laser scanner, an IMU and several digital cameras. The nominal localization accuracy of HiScan in GPS-denied environments can reach 2cm when aided by laser scanning control points (please note that, in this paper, the digital camera was not used). The performance parameters of MLS are listed in Table I.

The laser scanning control points were laid along the four sides of passageway in the indoor experimental zones, and distributed on the five sides of the teaching building. The normal coordinate values of the 20 laser scanning control points were established by using total station systems with higher accuracy (Leica TS15i, angle accuracy: 2″, distance accuracy: 3mm+1.5ppm). The MLS system was repeatedly driven four times both clockwise and counter clockwise around the passageway and teaching building at a constant speed. During the experiments, the IMU provided the direct reference from the MLS data with respect to the laser scanning control points. The commonly known estimated trajectory solution from the IMU system provides the position and orientation of the moving platform to convert the local coordinates of laser scanning point cloud into the world coordinate frame.

TABLE I
THE SPECIFICATIONS OF MLS

| Specification | Value |
| --- | --- |
| Laser Scanning Range | 0~200m |
| Output Data Rate | 976 000 pts/sec |
| Vertical FOV | 100° |
| Horizontal FOV | 300° |
| Laser Scanning Angle Accuracy | 0.01°rms |
| Laser Scanning Distance Accuracy | ±2mm |
| IMU Heading Accuracy | 0.012° |
| IMU Roll & Pitch accuracy | 0.008° |
| IMU Output Data Rate | 300 Hz |

In the indoor and outdoor scene, the sphere targets are used as the laser scanning control points, which are considered as the volumetric targets with 140 mm in diameter. The sphere target is made of high-strength PVC material which can allow the laser scanner obtaining the points data on spherical surface at farther distance. In indoor and outdoor experiments, the 20 laser scanning control points were selected to construct the 3D control network through observing the horizontal direction, the vertical angle, and the slope distance. The laser scanning control points were uniformly distributed in order to provide the sufficient observable access to the targets, to ensure an identical absolute position reference with the indoor control network. In indoor experimental environment, the four GPS reference stations were fixed as the known nominal points; they reached the 3D positional accuracy of 1cm in the geodetic coordinate frame. This accuracy level could provide the reliable reference for MLS system. The indoor network configuration along the passageway measurement is shown in Figure 3(a) and the outdoor network configuration along the teaching building is shown in Figure 3(b).

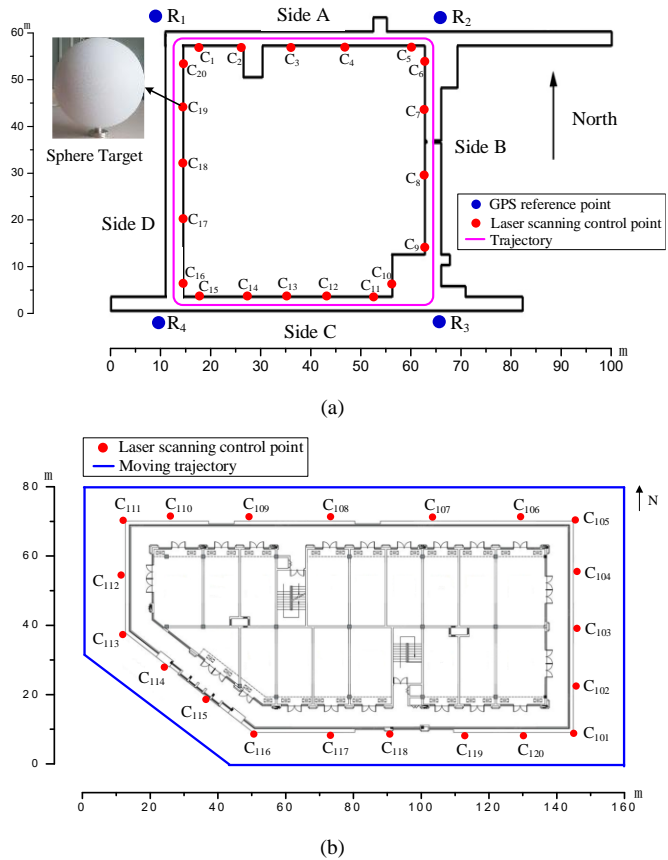

(a)

(b)

Fig. 3. Scatter plot of the 20 laser scanning control points and 4 GPS reference points: (a) indoor environment and (b) outdoor environment.

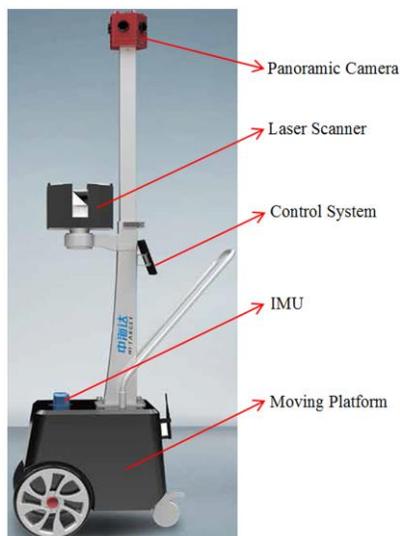

Fig. 4. Mobile laser scanning system

## V. RESULTS AND DISCUSSION

### A. Validation of LS, TLS and RWTLS-FIMLOE Methods for Control Position Accuracy

In order to validate the effectiveness of our proposed method, the positioning accuracy of the mobile laser scanning in the experimental test in Fig. (3) was examined. The performance of the proposed RWTLS-FIMLOE method was compared with that of existing popular algorithms in literature. As introduced in Section I, some popular algorithms such as LS, TLS, LS-SVM, NLS and LSC have been applied to improving the position accuracy and eliminate error sources for the MLS system in GPS-denied environments; however, the LS-SVM, NLS and LSC are subject to different modeling methods, detection distance and experimental environments. For this reason, in this study the RWTLS-FIMLOE method was only compared with LS and TLS in indoor and outdoor experimental environments. In addition, because RWTLS cannot calibrate the boresight alignment errors of MLS and FIMLOE is only a system parameter identifier, it is unnecessary to evaluate the positioning performance of individual RWTLS or FIMLOE.

In the comparative study, all 20 laser scanning control points were used in the experiments and a number of the overall quantitative indexes were calculated including the minimum, maximum, standard deviation and RMS (root mean square) indexes [36, 37]. As shown in the Figure 5, in indoor environment, RWTLS-FIMLOE method, TLS and LS method were used to estimate the position accuracy of the 20 laser scanning control points (The normal 3D coordinates of 20 control points had been established using high accuracy total station). The position errors Mean, Stdev and RMS of RWTLS-FIMLOE method in 2D (H) and 3D orientation are 1.36 cm, 0.94 cm, 1.37 cm and 1.85 cm, 0.95 cm, 1.86 cm, respectively. Similarly the TLS method in 2D (H) and 3D orientation are 2.00 cm, 1.03 cm, 2.02 cm and 2.47 cm, 1.12 cm, 2.50 cm, and the LS method in 2D (H) and 3D orientation are 2.36 cm, 1.06 cm, 2.38 cm and 3.00 cm, 1.16 cm, 3.03 cm, respectively.

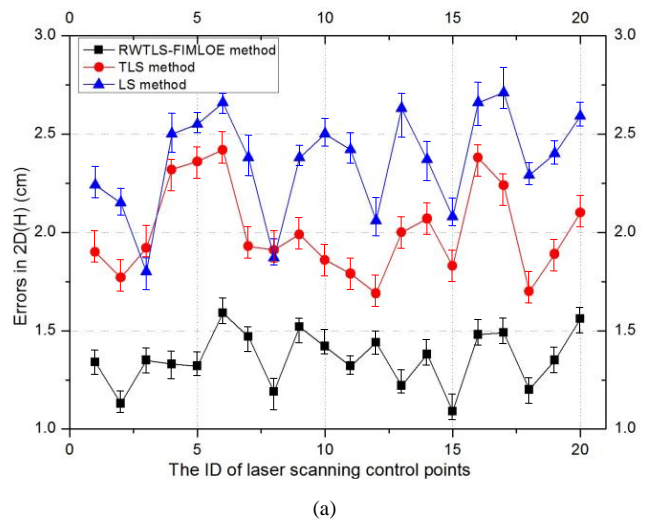

(a)

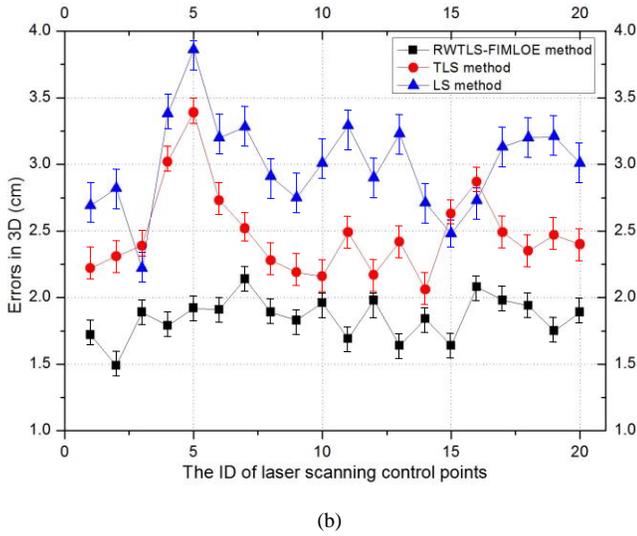

(b)

Fig. 5. The position errors of laser scanning control points using LS, TLS and RWTLS-FIMLOE method in indoor environments. (a) Errors in 2D(H) orientation; (b) Errors in 3D orientation.

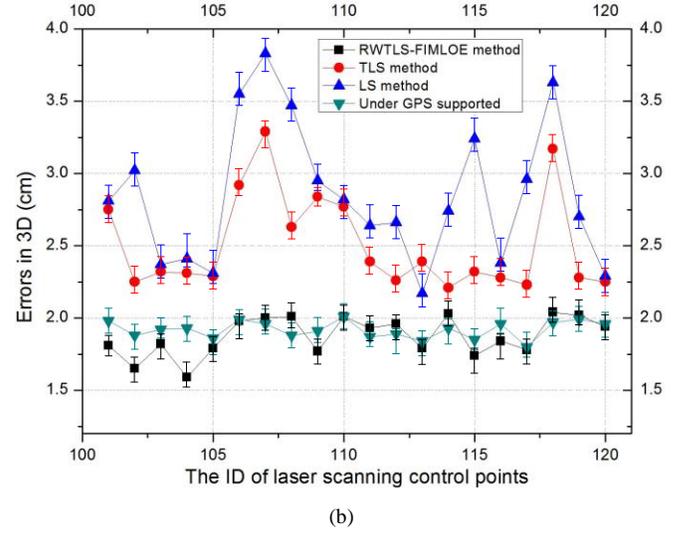

(b)

Fig. 6. The position errors of laser scanning control points using LS, TLS and RWTLS-FIMLOE method in outdoor environments. (a) Errors in 2D(H) orientation; (b) Errors in 3D orientation.

As shown in Figure 6, in outdoor environments the RWTLS-FIMLOE, TLS and LS method were used to estimate the position accuracy of the 20 control points, and their performance was compared with the position accuracy of MLS system with GPS supported. The measuring results are as follows: the position errors Mean, Stdev and RMS of MLS system under GPS supported in 2D (H) and 3D orientation are 1.66 cm, 0.85 cm, 1.67 cm and 1.91 cm, 0.86 cm, 1.92 cm, respectively. Similarly, the RWTLS-FIMLOE method are 1.52 cm, 0.92 cm, 1.53 cm and 1.87 cm, 0.93 cm, 1.88 cm, the TLS method are 2.10 cm, 1.08 cm, 2.12 cm and 2.50 cm, 1.13 cm, 2.53cm, and the LS method are 2.42 cm, 1.24 cm, 2.47 cm and 2.85 cm, 1.28 cm, 2.89 cm, respectively.

In summary, according to Figure 5 and 6, it is found that the proposed method can improve the MLS system position accuracy compared to the TLS and LS method, and it is suitable for GPS-denied environments.

### B. $\tau-test$ Method to Verify the Effectiveness of RWTLS-FIMLOE

To characterize the difference between the pre-surveyed coordinate and the adjusted MLS coordinate under indoor and outdoor environments, $\tau-test$ was employed to determine if the errors were significantly biased and if the errors were much larger than what the RWTLS-FIMLOE method has required.

Specifically, a $\tau-test$ was built up to test if a mean error value $m$ was significantly different from zero under the null hypothesis: $H_0: m=0$ vs. the alternate hypothesis $H_a: m \neq 0$:

$$\tau = \frac{\hat{m}}{\hat{\sigma}_{\hat{m}}} = \frac{\hat{m}}{\hat{\sigma}}\sqrt{n} \sim \tau_{n-1} \qquad (26)$$

where $\hat{m}$ is the mean value of a group error samples; $\hat{\sigma}_{\hat{m}} = \hat{\sigma}/\sqrt{n}$ is the standard deviation of mean value $\hat{m}$, $\hat{\sigma}$ is the standard deviation of a group error samples; $n$ is the number of the samples; $\tau_{n-1}$ is a $\tau-test$ with the degrees of freedom of $n-1$.

Furthermore, a $\chi^2$ was constructed to statistically conclude if a sample standard deviation was satisfied with the specifically required accuracy level under the hypothesis.

$$\begin{cases} H_0: \sigma^2 = \sigma_0^2 \\ H_a: \sigma^2 > \sigma_0^2 \, or \, \sigma^2 \neq \sigma_0^2 \end{cases} \qquad (27)$$

The $\chi^2$ test value was denoted as

$$\chi^2 = \frac{\hat{\sigma}^2}{\sigma_0^2}(n-1) \sim \chi_{n-1}^2 \qquad (28)$$

where $\sigma_0$ is a given standard deviation that indicates a required accuracy level, and $\chi_{n-1}^2$ is the Chi-square test with $(n-1)$ degrees of freedom.

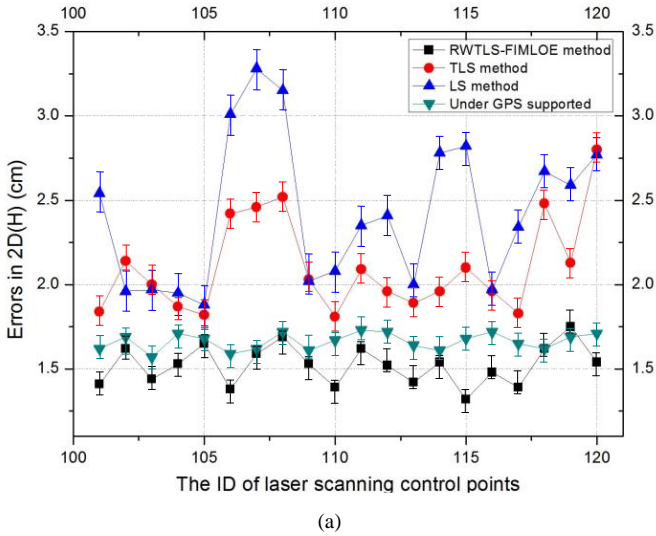

(a)

TABLE II.
THE ANALYSIS OF THE DIFFERENCE BETWEEN LS, TLS AND RWTLS-FIMLOE METHODS USING $\tau-test$

| Method | Error | Mean (cm) | Stdev (cm) | RMS (cm) | $N_C$ | Two-tailed $\tau$-test α=0.05% $H_0: \mu = 0$ | | $\chi^2$ Test(α=0.05%)vs the 95%accuracy | | | |
|---|---|---|---|---|---|---|---|---|---|---|---|
| | | | | | | | | $H_0: \sigma^2 = 1.5^2 cm^2$ | | $H_1: \sigma^2 = \sigma_a^2$ | |
| | Error (cm) | Mean (cm) | Stdev (cm) | RMS (cm) | $N_C$ | $\tau$ | $\tau_{C0.025}$ | $\chi^2$ | $\chi^2_{C0.05}$ | $\chi^2$ | $\sigma_a$ (cm) |
| LS | N | -1.80 | 2.03 | 2.65 | 20 | -3.31 | 2.16 | | | | |
| | E | 1.48 | 1.72 | 2.22 | 20 | 3.21 | 2.16 | | | | |
| | U | -1.25 | 1.67 | 2.04 | 20 | -2.81 | 2.16 | 61.73 | 22.36 | 22.22 | **2.52** |
| | 2D(H) | 2.33 | 2.66 | 3.46 | 20 | | | 163.63 | 22.36 | 21.90 | **4.13** |
| TLS | N | -0.62 | 1.21 | 1.32 | 20 | -1.92 | 2.16 | | | | |
| | E | 0.65 | 1.38 | 1.48 | 20 | 1.77 | 2.16 | | | | |
| | U | 0.68 | 1.31 | 1.43 | 20 | 1.93 | 2.16 | 38.07 | 22.36 | 21.41 | **2.03** |
| | 2D(H) | 0.90 | 1.83 | 1.98 | 20 | | | 77.69 | 22.36 | 22.30 | **2.84** |
| RWTLS -FIMLOE | N | -0.46 | 0.99 | 1.06 | 20 | -1.73 | 2.16 | | | | |
| | E | -0.48 | 1.13 | 1.19 | 20 | -1.58 | 2.16 | | | | |
| | U | -0.34 | 1.19 | 1.20 | 20 | -1.06 | 2.16 | 31.57 | 22.36 | 21.93 | **1.82** |
| | 2D(H) | 0.66 | 1.51 | 1.59 | 20 | | | 52.41 | 22.36 | 22.29 | **2.33** |

TABLE III.
THE EFFECT OF LASER SCANNING CONTROL POINT'S NUMBER ON MLS POSITION ACCURACY BASED ON RWTLS-FIMLOE METHOD

| Results | With 20 control points (cm) | | | | | With 12 control points (cm) | | | | | With 8 control points (cm) | | | | |
|---|---|---|---|---|---|---|---|---|---|---|---|---|---|---|---|
| | N | E | U | 2D | 3D | N | E | U | 2D | 3D | N | E | U | 2D | 3D |
| Min | -2.03 | -1.83 | -1.96 | 1.04 | 1.42 | -1.88 | -2.27 | -2.87 | 1.29 | 1.56 | -2.94 | -2.71 | -2.87 | 1.52 | 2.87 |
| Max | 1.33 | 1.51 | 1.40 | 2.53 | 2.76 | 1.56 | 1.25 | 1.48 | 2.75 | 3.74 | 1.56 | 1.78 | 3.33 | 3.46 | 4.80 |
| Mean | -0.57 | -0.68 | -0.21 | **0.89** | **0.91** | -0.57 | -0.80 | -0.67 | **0.98** | **1.19** | -1.56 | -1.15 | -1.65 | **1.94** | **2.54** |
| RMS | 1.21 | 1.16 | 1.20 | **1.67** | **2.06** | 1.43 | 1.28 | 1.51 | **1.92** | **2.45** | 2.07 | 1.79 | 2.32 | **2.74** | **3.58** |
| Stdev | 1.12 | 0.98 | 1.24 | **1.49** | **1.93** | 1.38 | 1.05 | 1.43 | **1.73** | **2.24** | 1.42 | 1.44 | 1.71 | **2.02** | **2.65** |

It is supposed that the accuracy evaluation requirement of the indoor localization is 2cm horizontal and 3cm vertical accuracy criterion at the 95% confidence level. If the test in Equation (28) is rejected at a specific error level of Type I Error, then a different alternate value $\sigma_a^2$ can be chosen as a substitution of $\sigma_0^2$ to find out the lower bound that can pass the test in Equation (28). This lower bound is called the achieved accuracy with the involved samples.

Similarly, this test statistics can also be applied to the RMS value of a group of samples. Supposing that all the tests in the accuracy assessment in this paper were performed at the 5% significance level of Type I Error. Table II presents a summary of $\tau-test$ and alternate $\chi^2$ test of the differences between the pre-surveyed coordinates and the original MLS system coordinates using LS, TLS and RWTLS-FIMLOE methods. The achieved accuracies were ±2cm (horizontal) and ±3cm (vertical) at the 95% confidence level according to the alternate $\chi^2$ test. Using RWTLS-FIMLOE method, the horizontal and the vertical localization accuracy have achieved ±1.82cm and ±2.33cm, respectively, the horizontal and the vertical localization accuracy of TLS and LS method are ± 2.03cm, ±2.84cm and ± 2.52cm, ±4.13cm, respectively. Compared with TLS and LS method, RWTLS-FIMLOE method can improve the overall accuracy by 17.96%, 10.35% and 43.58%, 27.28% in horizontal and vertical localization.

The comparisons between the three methods indicated that the RWTLS-FIMLOE method using control points from all side is able to achieve better accuracy in both horizontal and vertical directions compared to LS and TLS methods.

*C. The impact of the Number of Laser Scanning Control Points on RWTLS-FIMLOE Method*

The main purpose of this test is to investigate the effect of the number of the laser scanning control points on RWTLS-FIMLOE method. There were a total of 20 laser scanning control points in the experimental scene, the position accuracy improving of MLS system solutions based on the RWTLS-FIMLOE methods were compared through the laser scanning control points using different number of control points (20, 12 and 8) incorporating with the same feature constraints. Moreover, the feature constraints were sequentially tested using $\tau-test$ to detect any inconsistency with the precious observation group, and only the qualified constrains were used in the RWTLS-FIMLOE method. The plot of the selected control points in the experimental scene is presented in Figure 7. The experimental results are shown in Table III, the correction effectiveness of the mean errors had achieved the best accuracy which is 0.89 cm in 2D(H) and 0.91 cm in 3D by using 20 control points. Moreover, the 2D(H) and 3D orientation accuracies could achieve 0.98 cm and 1.19 cm when only 10 control points were applied. Although using more control points could achieve better accuracy, the positioning performance would not be decreased significantly by reducing the number of control points down to 60 percentage of the total number of the control points. Furthermore, for the test where only 8 control points were used, the accuracy became 1.94cm in 2D(H) and 2.54cm in 3D, which were decreased relatively significant in comparison with the above two cases. Based on the results among all test cases, using more control points in the solution refining process could generally achieve better accuracy in both 2D(H) and in 3D. However, the overall accuracies were not decreased significantly if reducing the number of the used control points up to 60 percent of the total number. However, the 3D positioning accuracy was decreased dramatically using only 40 percent of total number in the 3D position. Therefore, it is necessary to select the control points efficiently and effectively by considering the project budget and the minimal required accuracy.

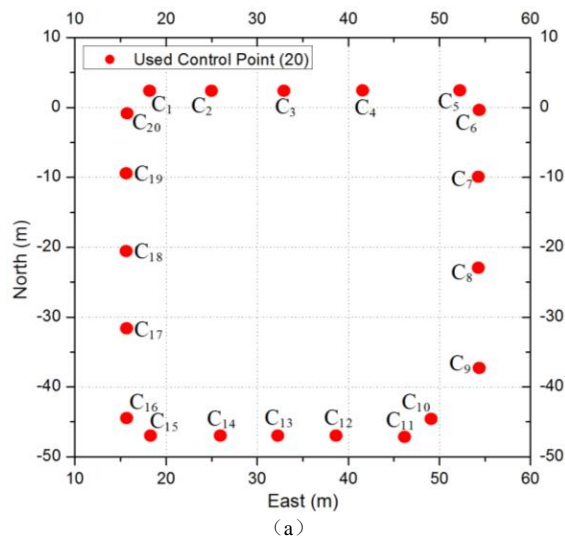

(a)

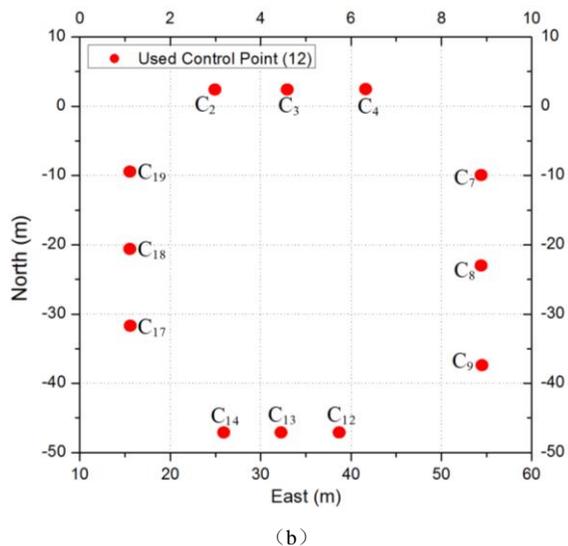

(b)

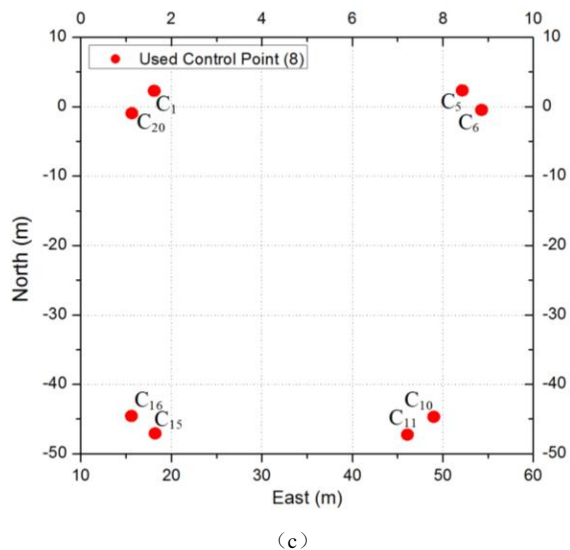

(c)

Fig. 7 The scatter plot of coordinates in 2D horizontal of experimental scene. (a) 20 used control points, (b) 12 used control points, (c) 8 used control points.

## VI. CONCLUSIONS

This paper proposed a novel algorithm which integrates RWTLS and FIMLOE methods to improving the positioning accuracy of the MLS system in GPS-denied environment. This new method inherits the advantages of RWTLS and FIMLOE methods. The primary contributions of this paper are:

(1) An integration method which combines the advantages of RWTLS and FIMLOE is proposed. This new method is able to correct the 3D point observation model and the uncertainty propagation parameters vector, and it is also capable of calibrating the boresight alignment errors between the laser scanner frame and the IMU frame.

(2) This paper established a mathematical model for MLS and analyzed in depth the effects of the individual error source on the error budget of MLS.

(3) The experimental results show that the proposed method can improve the positioning accuracy of the MLS system in terms of Mean, RMS and Stdev in the 2D(H) and 3D orientations compared with TLS and LS methods, , and it is suitable for GPS-denied environments. Furthermore, according to the alternate $\chi^2$ test, the RWTLS-FIMLOE method can bring improvements in the overall accuracy up to 17.96%, 10.35% and 43.58%, 27.28% in horizontal and vertical localization when compared with TLS and LS methods.

In conclusion, the developed RWTLS-FIMLOE method realized an effective improvement for MLS system position accuracy and exhibited similar performance of TLS. It can be widely applied in MLS surveying engineering, such as 3D indoor modeling for visualization in planning, simulations for environmental management, navigation for robot or vehicle, and surveying for underground mine, and so on. It is able to keep the high accuracy of the laser scanning points cloud data with little uncertainty in GPS-denied environments.

ACKNOWLEDGMENT

We would like to thanks LetPub (www.letpub.com) for providing linguistic assistance during the preparation of this manuscript.